\documentclass[twocolumn,aps,pra]{revtex4}
\usepackage{psfig}
\usepackage{epsfig}

\begin{document}
\title{Healing Length and Bubble Formation in DNA}

\author{Z. Rapti$^{1}$, A. Smerzi$^{2,3}$,  K. \O. Rasmussen$^{2}$ and A. R. Bishop$^{2}$ }
\affiliation{$^1$ Center for Nonlinear Studies, Los Alamos National Laboratory, Los Alamos, 
New Mexico 87545, and Department of Mathematics, University of Illinois at Urbana-Champaign, 
1409 W. Green Street, Urbana, IL 61801}
\affiliation{$^2$ Theoretical Division and Center for Nonlinear Studies,
Los Alamos National Laboratory, Los Alamos, New Mexico 87545}
\affiliation{$^3$ Istituto Nazionale di Fisica per la Materia BEC-CRS,
Universit\`a di Trento, I-38050 Povo, Italy} 
\author{C.H. Choi, A. Usheva }
\affiliation{Endocrinology, Beth Israel Deaconess Medical Center and Harvard Medical School,
Department of Medicine, 99 Brookline Avenue, Boston, Massachusetts 02215}

\date{\today}

\begin{abstract}
We have recently suggested that the probability for the formation 
of thermally activated DNA bubbles  
is, to a very good approximation, proportional to the number 
of soft AT pairs over a length $L(n)$ that depend on the size $n$ 
of the bubble 
and on the temperature of the DNA.
Here we clarify the physical interpretation of this length by 
relating it to the (healing) length that is required for 
the effect of a base-pair defect to become neligible.
This provides a simple criteria to calculate $L(n)$ for
bubbles of arbitrary size and for any temperature of the DNA.
We verify our findings by exact calculations of the equilibrium
statistical properties of the Peyrard-Bishop-Dauxois model.
Our method permits calculations of equilibrium thermal openings 
with several order of magnitude less numerical expense as compared
with direct evaluations.
\end{abstract}
 
\maketitle

\section {Introduction.}

Local separation of double-stranded DNA into single-stranded
DNA is fundamental to transcription and other important
intra-cellular processes in living organisms. In equilibrium,
DNA will locally denaturate when the free energy of the separated single-stranded DNA
is less than that of the double-stranded DNA. Because of the larger entropy of the flexible
single-strand, the more rigid double-strand can be thermally destabilized locally to form temporary
``bubbles'' in the molecule already at physiological temperatures \cite{stjaalet}.
Considering this entropic effect together with the 
inherent energetic heterogeneity --
GC base pairs are 25 \% more strongly bound than the AT bases --
of a DNA sequence, it is plausible that certain regions (subsequences) are more 
prone to such thermal destabilization than others. In fact recent work \cite{donald} demonstrates not only that 
such a phenomena exists but more importantly that the location of these large bubble openings in a variety of 
DNA sequences coincide with sites active during transcription events. This discovery represent a significant
advance in the understanding of the relationship between local conformation and function in 
bio-molecules. While there is no guarantee that this mechanism applies to all transcription 
initiation events, the present agreement is very encouraging. Similarly, other large bubbles identified 
may well have a relationship to other biological functions. The agreement is based on the Peyrard-Bishop-Dauxois (PBD) model, 
\cite{dpb}, which evidently contains some essential basic ingredients --local constraints (nonlinearity), base-pair 
sequence (colored disorder) and entropy (temperature). 
The equilibrium thermodynamic properties of the model were numerically calculated from the
partition function using the
transfer integral operator method (TIO). (A complementary 
direct numerical evaluation of the partition function has been reported in Ref. \cite{peynew}). This allows 
the precise evaluation of
probabilities of bubbles as a function of temperature, location in a given base-pair sequence, and 
bubble size. 
In recent work \cite{euro}, we reported that the probabilities of finding bubbles 
extending over $n$ sites do not depend on a specific DNA subsequences. Rather, such probabilities depend on 
the density of 
soft A/T base pairs within a region of length $L(n)$. Here we suggest that 
this characteristic length is simply related to the characteristic distance away from an AT base pair -- considered as a 
defect placed in a homogeneous GC-sequence-- where the probability values of the base pairs return 
to the GC bulk-value.
Lastly, based on this concept of effective density approximation, we examine 
five different human promoter sequences, and demonstrate the striking agreement in the predictions from the two methods.  

\section {The PBD model and the TIO method.}
The potential energy of the PBD model is \begin{eqnarray}
{E}=\displaystyle \sum_{n=1}^{N} \left[
V(y_n)+W(y_n,y_{n-1})\right] = \sum_{n=1}^N {\cal E}(y_n,y_{n-1})\cr.
\label{ham}
\end{eqnarray}
Here $V(y_n)=D_n (e^{-a_n y_n}-1)^2$, represents the nonlinear hydrogen bonds between the bases;
$W(y_n,y_{n-1})=\frac{k}{2}\left(1+\rho e^{-b(y_n+y_{n-1})}\right)(y_n-y_{n-1})^2$ is the nearest-neighbor 
coupling that represents the (nonlinear) stacking interaction
between adjacent base pairs: it is comprised of a harmonic coupling with a state-dependent coupling constant 
effectively modeling the change in stiffness as the double strand is opened (i.e. entropic effects).
The sum in Eq.(\ref{ham}) is over all base-pairs of the molecule and $y_n$ 
denotes the relative
displacement from equilibrium bases at the $n^{th}$ base pair. The importance of the heterogeneity of the 
sequence is 
incorporated by assigning different values to the parameters of the Morse potential, depending on the 
the base-pair type.
The parameter values we have used are those in Refs. \cite{campa, para} chosen to reproduce a variety 
of experimentally observed thermodynamic properties.

The equilibrium thermodynamic properties of the PBD model can be calculated from
the partition function
\begin{eqnarray}
{\cal Z} &=&\int \prod_{n=1}^N dy_n e^{-\beta{\cal E}(y_n,y_{n-1})} \label{parf}\cr 
&=&\int \prod_{n=s}^{s+k-1} dy_nZ_k(s)\,e^{-\beta{\cal E}(y_n,y_{n-1})},
\end{eqnarray}
where we have introduced the notation
$$Z_{k}(s)=\int \prod_{n \ne s,...,s+k-1}^N  dy_n\, e^{-\beta {\cal E}(y_n,y_{n-1})}$$ 
and $\beta= (k_BT)^{-1}$ is the Boltzmann factor.
In order to evaluate the partition function (\ref{parf}) using the TIO method, we first symmetrize 
$e^{-\beta {\cal E}(x,y)}$ by introducing \cite{dp} 
\begin{eqnarray}
S(x,y)&=&\exp\left(-\frac{\beta}{2} (V(x)+V(y)+2 W(x,y))\right) \label{kernel}\cr
      &=&S(y,x). \nonumber
\label{symker}
\end{eqnarray}
Here the second equality holds only when $x$ and $y$ correspond to base-pairs of the same kind.
Using Eq. (\ref{parf}) the expression for $Z_k(s)$ is rewritten as 
\begin{eqnarray}
Z_k(s)=&&\int \left(\displaystyle \prod_{n \ne s,...,s+k-1}^N dy_n S(y_n,y_{n-1}) \right)\cr
&&\times dy_0 e^{ -\frac{\beta}{2} V(y_1)} e^{ -\frac{\beta}{2} V(y_N)},
\label{sparf}
\end{eqnarray}
where open boundary conditions at $n=1$, and $n=N$ have been used.
To proceed, a Fredholm integral equations with a real symmetric kernel
\begin{eqnarray}
\int dy S(x,y) \phi(y)=\lambda \phi(x)
\label{intsym}
\end{eqnarray}
must be solved separately for the A/T and for the G/C base-pairs. 

Since the eigenvalues are orthonormal and the eigenfunctions form a complete basis, Eq.(\ref{intsym}) can 
be used sequentially to replace all integrals by matrix multiplications in Eq. (\ref{sparf}). 
Unlike in Ref. \cite{zhang} where the kernels $S(x,y)$ were expanded in terms of orthonormal bases, here we 
choose to use Eq. (\ref{intsym}) iteratively. In this way we reduce the number of integral equations that need 
to be solved from 4 to 2, and at the same time the matrices that need to be multiplied are lower dimensional.
Whenever the sequence heterogeneity results in a non-symmetric $S(x,y)$, Eq.(\ref{intsym}) cannot be
used and we resort to a symmetrization technique, based on successive introduction of auxiliary integration 
variables, as explained in Ref. \cite{thesis}. 

We evaluate the probabilities $P_{k}(s)$, for a base-pair opening spanning k base-pairs (our operational
definition of a bubble of size $k$), starting at base-pair $s$ as
\begin{equation}
P_{k}(s)={\cal Z}^{-1}{\int_t^{\infty}\prod_{n=s}^{s+k-1} dy_n Z_k(s)e^{-\beta{\cal E}(y_n,y_{n-1})}}
\label{prob}
\end{equation}
where $t$ is the separation (which we have here taken as 1.5 {\AA}) of the double strand above which we define 
the strand to be melted. 

\section {Lengthscales and effective density approximation.}
In Ref.\cite{euro} we suggested that
the probabilities of finding bubbles 
extending over $n$ sites localized around a given bp, is, to
a very good approximation, proportional to the 
density of soft A/T base pairs within a region of length $L(n)$ 
centered around the same bp,
an approach we term here as effective density approximation (EDA).
 The lengths $L(n)$ were obtained from numerical
transfer integral calculations of the 
bubble probabilities of several simple (but experimentally
realizable) sequences. The A/T density profiles were
therefore compared with
the exact probabilities for thermal activation of bubbles of
sizes $n=1$ and $n=5$ of a wild and a mutant version of the AAV P5
promoter. The agreement was excellent. However, no physical explanation
for the origin of these characteristic lengths was provided nor were they connected to any intrinsic 
length of the PBD model. But, since they appear prominently in the formation of DNA bubbles, it is important 
to investigate both of these questions.

\begin{figure}[h]
\centerline{\psfig{figure=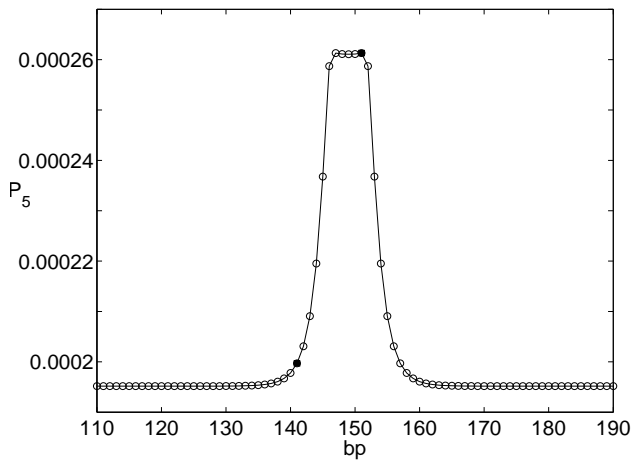,width=83mm,height=65mm}}
\caption{The probability profile for the creation of a bubble of size $5$ bp.
The isolated A/T bp embedded in a sequence of G/C bps at $bp=151$ is denoted by a solid black circle.  
A second black circle is located at $bp=151-L(5)=141$. The relative error 
$\frac{P_5(141)-P_5(110)}{P_5(110)}=0.0232$.}
\label{p2}
\end{figure}

\begin{figure}[h]
\centerline{\psfig{figure=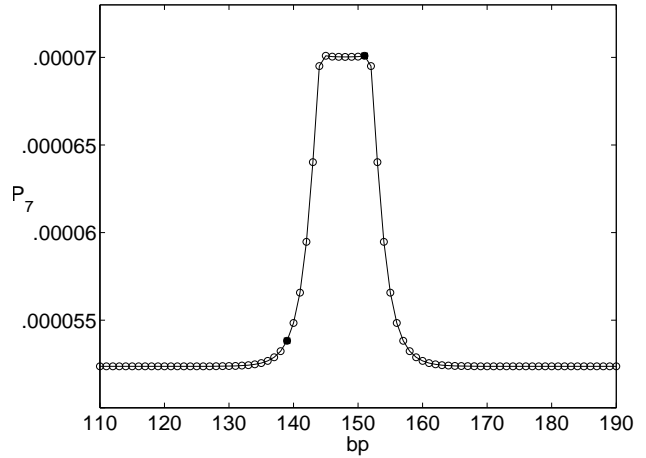,width=83mm,height=65mm}}
\caption{The probability for the creation of a bubble of size $7$ bp.
The black circle at $bp=151$ represents the defect. The second black circle is located 
at $bp=151-L(7)=139$. The relative error $\frac{P_7(139)-P_7(110)}{P_7(110)}=0.0279$.}
\label{p3}
\end{figure}

\begin{figure}[h]
\centerline{\psfig{figure=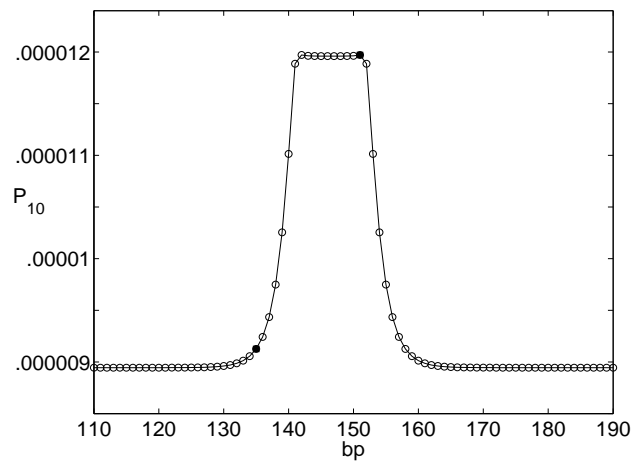,width=83mm,height=65mm}}
\caption{The probability for the creation of a bubble of size $10$ bp.
As before, the defect is represented by a solid black circle, and a second one is  located at 
$bp=151-L(10)=135$. The relative error $\frac{P_{10}(135)-P_{10}(110)}{P(110)}= 0.0203 $.}
\label{p4}
\end{figure}

In Figs.\ref{p2}-\ref{p4} we consider a sequence composed of $150$ G/C $+\,1$ A/T $+\, 150$ G/C.
In other words, we place a defect (A/T instead of G/C) at
the site $l_0=151$. This defect is $150$ bp away from the two ends of the sequence in order to eliminate 
boundary effects.

A/T base pairs have a smaller bonding energy than GC bps. Therefore, the A/T defect 
softens a number of GC-bps around it and increases the opening probability. 
Clearly, sufficiently away from the defect the opening
probability regains the bulk value of a homogeneous G/C-sequence at the given temperature, given threshold 
and given bubble size. Our claim is that the characteristic length
$L(n)$ is the distance necessary
to be away from the defect so that the G/C bps there are no longer affected. 
This can be quantified by calculating the relative fluctuation 
\begin{equation}
{\frac{P_n(l_0-L(n))-P_n(110)}{P_n(110)}}=\alpha,
\label{alpha}
\end{equation}
where $l_0-L(n)$ is
the bp site obtained counting $L(n)$ downstream from the defect site, see Figs.\ref{p2}-\ref{p4}, and
at site $110$ we assume that the bulk value has been regained.
The remarkable finding is that with the choice of $L(n)$ considered in our previous work \cite{euro},
obtained independently by merely fitting the full numerical TIO calculations of the bubble formation probabilities
of different simple sequences,
we obtain from Eq.(\ref{alpha}) $\alpha \simeq 2.5 \%$, independently from the size of the bubble and the
temperature of the DNA sequence. This can be seen in Figs.\ref{p2}-\ref{p4}: the circle at $bp=151=l_0$ is the
A/T defect, while the circles at $bp=141,139,135$ are the positions of the bp at $l_0-L(n)$. 
We can therefore reverse the perspective and {\it define} the characteristic length as the one
given by Eq.(\ref{alpha}), with $\alpha \simeq 2.5 \%$. This is important for pratical applications, 
since it gives a simple criterion to estimate bubbles probabilities for arbitrary bubble sizes and DNA 
temperatures (and arbitrary PBD inter base-pairs interaction parameters), but it also immediately  
suggests a simple physical explanation for $L(n)$.

We parametrize 
the decay of the probability values as a function of the downstream
distance from the A/T defect 
according to
\begin{equation}
P_n(l)=A_n+B_n\,exp [{-\frac{l_0-n+1-l}{\xi_n}}], 
\label{prob_decay}
\end{equation}
where $P_n(l)$ is the probability for finding a bubble of size $n$ located at the site
$l$, $A_n$ is the bulk value of the homogeneous G/C sequence
and $A_n + B_n$ is the value of the probability at the site $l_0-n+1$, which is the same as the
probability value of the defect site $l_0$.
$\xi_n$ is the healing length of the system, namely the characteristic length for the 
perturbation to die out, 
which, quite generally, depends on the size $n$ of the bubble, temperature of the DNA and the 
parameters of the PBD model. Replacing Eq.(\ref{prob_decay}) in Eq.(\ref{alpha}), we obtain the
relation $L(n) = n -1 + \xi_n~ \ln\left(\frac{B_n}{\alpha\,A_n}\right)$, where $B_n / A_n = 
(P_n(l_0)-P_n(bulk))/P_n(bulk)
\simeq 0.34$. We emphasize that both $B_n/A_n$ and $\alpha$ are independent
of the size $n$ of the bubble.
It follows that there is a simple linear relation between the healing and characteristic lengths: 
\begin{equation}
L(n)=n-1+2.6\,\xi_n.
\label{relation}
\end{equation}
This is a very important result of this report. The healing length
can be easily calculated as a function of the bubble size and temperature with an homogeneous G/C sequence plus a 
single defect, as shown above. From this, we can calculate the value of $L(n)$ and,
therefore, estimate the probability for the creation of bubbles for 
arbitrary DNA sequences at any temperature.
For instance, for bubbles of size $n=7$, we obtain $L(7) = 12$, while
for bubbles of size $n=10$ we have $L(10)=16$ at $T=300~K$ and PBD
parameters as in \cite{para}. 

In order to examine how the values of the parameters of the PBD model affect those of $L(n)$, 
we set $\rho=0$ and repeat the calculation of $L(10)$. Since, when $\rho$ decreases, so does the 
"cooperativity" of the base base pairs, one would expect to observe a drop in the $L(10)$
value. This is indeed the case: $L(10)=14$, while for $\rho=2$ the value as 16.
\begin{figure}[h]                                                                                                      
\centerline{\psfig{figure=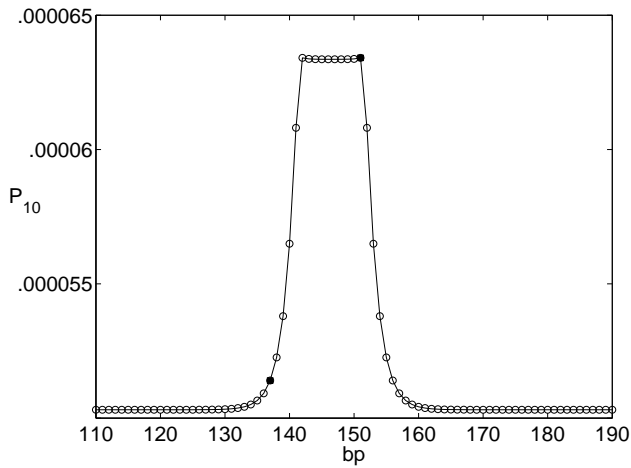,width=83mm,height=65mm}}                                                        
\caption{In this figure we show the probability for the creation of a bubble of size $10$ bp,
when the coupling constant $\rho=0$. As is indicated by the relative error 
$\frac{P_{10}(137)-P_{10}(110)}{P(110)}= 0.0216 $, now $L(10)=14$.}                             
\label{p5}                                                                                                             
\end{figure}  
We will show in the next Section 
how this approach compares with exact transfer integral operator
calculations of the statistical properties of the PBD model. 

We conclude this Section by noting that the Figs. \ref{p2}-\ref{p4}
exhibit symmetry, but not with respect to the defect. While the defect is always at $bp=151$,
the symmetry is with respect to bp 149 in the $P_5$ case, 148 in the
$P_7$ case, and the axis that separates bps 146-147 in the $P_{10}$ case.
Another feature is the existence of a second local maximum with the
same value as $P_n(151)$, and a slight drop in the probability values in the
middle of the peak. 
We notice that the two maxima are located at sites $l_0$ and $l_0-n+1$. 
This suggests that a bubble with a defect at its boundary has a higher
probability to form: in the $P_n(l_0 - n +1)$ case the defect is at the
end of the bubble, while in the $P_n(l_0)$ case it is at the begining of the bubble. 
Also, the probability drops in the middle of the peak because the bubble there contains a
defect that is trapped within G/C bps, and it turns out that the
probability of formation of a bubble of this kind is smaller.

\section {Comparison of the EDA and TIO Method.} 
We now compare the probability profiles obtained from the effective density approach with the characteristic
length $L(n)$ calculated as in the previous Section, with exact results obtained with
the TIO method. We consider five different human genome subsequences, and compare the calculations
for the probability of formation of bubbles of sizes $n=7$ and $n=10$.

\begin{figure}[h]
\centerline{\psfig{figure=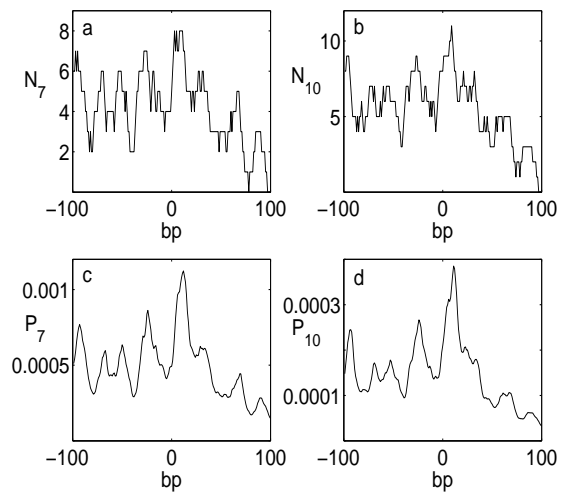,width=73mm,height=65mm}}
\caption{Effective density profiles for $7$ and $10$-site long bubbles
 (a,b) and probability profiles calculated with the transfer
  integral approach, (c,d). The sequence is the cox 8 promoter.}
\label{cox8}
\end{figure}

\begin{figure}[h]
\centerline{\psfig{figure=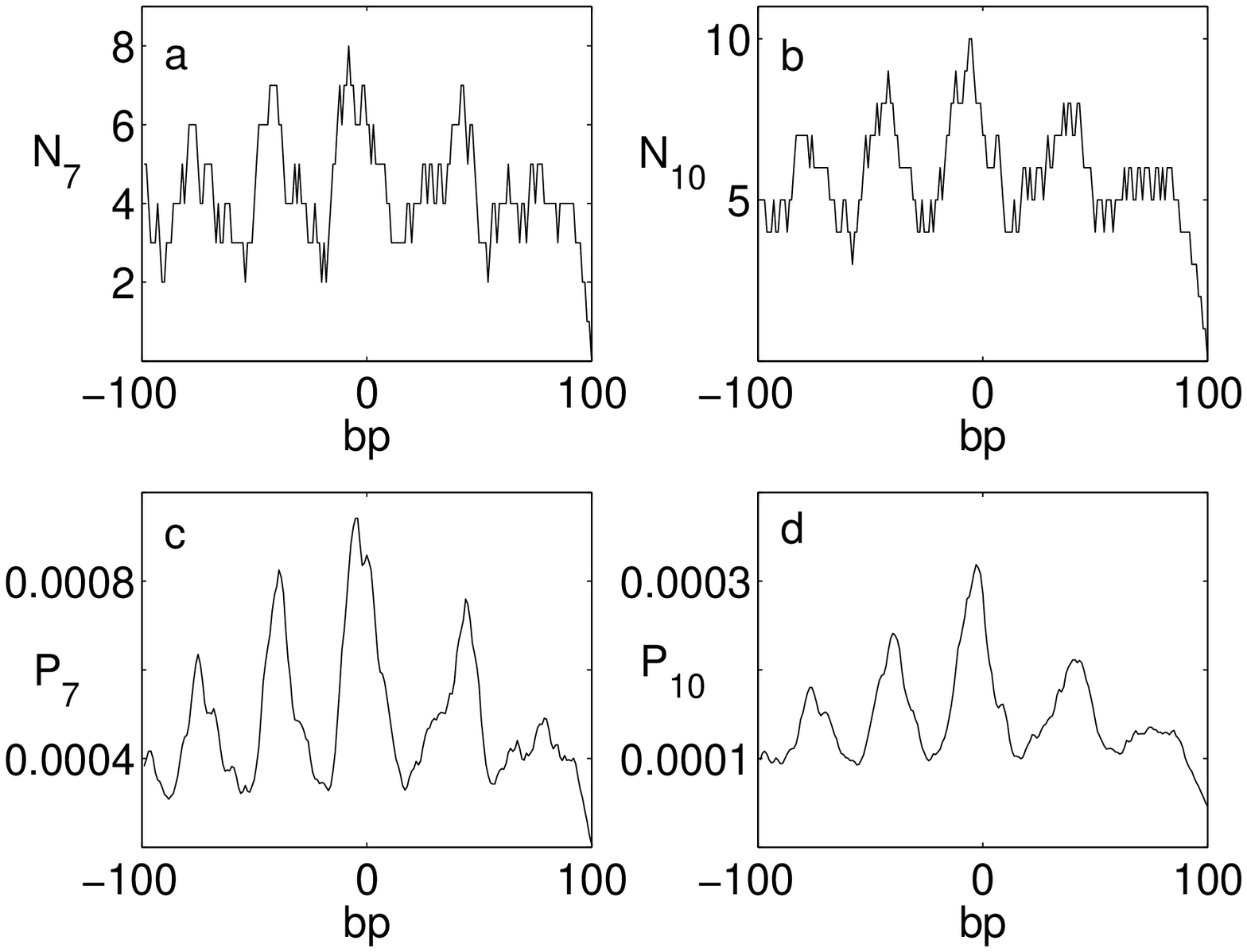,width=73mm,height=65mm}}
\caption{Effective density profiles for $7$ and $10$-site long bubbles
 (a,b) and probability profiles calculated with the transfer
  integral approach, (c,d). The sequence is the cox 11 promoter.}
\label{cox11}
\end{figure}

\begin{figure}[h]
\centerline{\psfig{figure=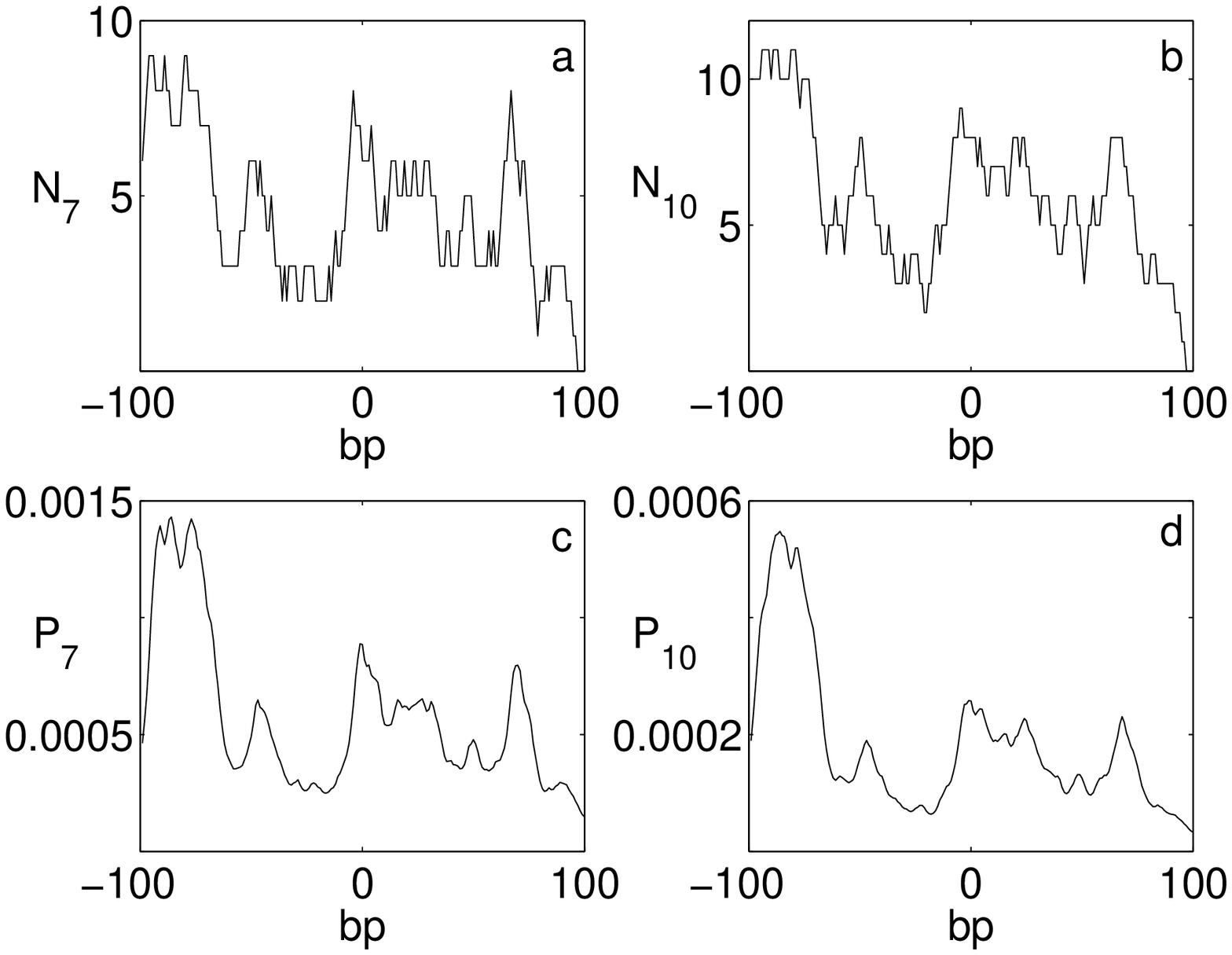,width=73mm,height=65mm}}
\caption{Effective density profiles for $7$ and $10$-site long bubbles
 (a,b) and probability profiles calculated with the transfer
  integral approach (c,d). The sequence is the gtf2f2 promoter.}
\label{gtf2f2}
\end{figure}

\begin{figure}[h]
\centerline{\psfig{figure=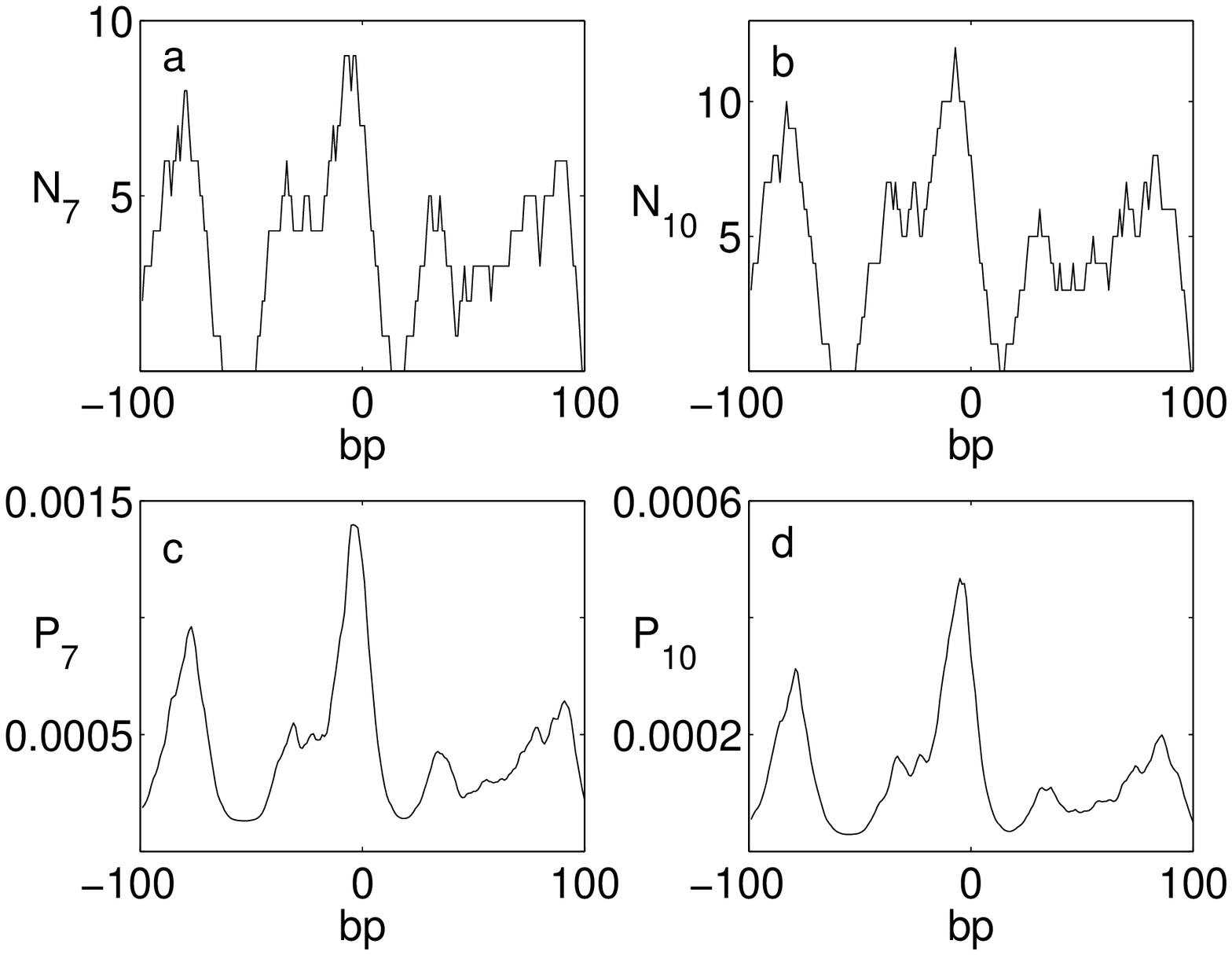,width=73mm,height=65mm}}
\caption{Effective density profiles for $7$ and $10$-site long bubbles
 (a,b) and probability profiles calculated with the transfer
  integral approach (c,d). The sequence is the h33a promoter.}
\label{h33a}
\end{figure}

\begin{figure}[h]
\centerline{\psfig{figure=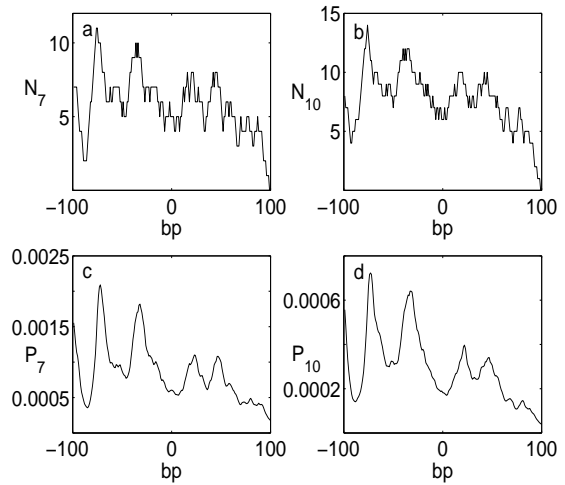,width=73mm,height=65mm}}
\caption{Effective density profiles for $7$ and $10$-site long bubbles
 (a,b) and probability profiles calculated with the transfer
  integral approach (c,d). The sequence is the h3b promoter.}
\label{h3b}
\end{figure}

In the panels $(a,b)$ of Figs.\ref{cox8}-\ref{h3b} we plot (as a function of the 
bp site) the number $N_7$ and $N_{10}$ of A/T bps calculated over a distance $L(7)=12$ (panel a) and 
$L(10)=16$ (panel b). These A/T density profiles can be compared with the probability for the thermal 
creation of bubbles of seven, $P_7$, and ten, $P_{10}$, sites, panels $(c,d)$. In all cases (and in several 
other not reported here) the resemblance in the main features of the respective profiles is striking.
In particular, EDA correctly predicts the locations and relative 
weights of the probability peaks. The crucial point is that, while the profiles obtained with 
the EDA requires few seconds to be calculated, the full TIO methods is very
time consuming (of the order of several hours in the cases presented here).
To fully appreciate this advantage, we note that with the EDA the entire human genome can be 
sequenced for bubble formation probabilities in few minutes, while a statistical approach based on the
calculation of the partition function is clearly impossible. 
\section {Conclusions.}
It has been suggested that the DNA transcription initiation sites can coincide with
the location of large bubble openings. A thorough
investigation of this hypothesis requires the statistical analysis of
many DNA promoters within the PBD model. Such a task becomes  
quickly prohibitive when studying bubble-promoter correlations in a
significantly 
large number of cases (namely, for large sequences). This problem has 
motivated the development of an alternative simplified approach to
calculate the bubble formation probabilities. We have found that this
probabilities are
proportional to the density of soft A/T base-pairs calculated over an
effective length which depends on the size of the bubble and the
temperature of the DNA. We have clarified the physical origin
of such a length and suggested a simple procedure for its caluclations. 
The results of our effective density approach are in extremely good
agreement with full exact calculations, but with a
numerical effort reduced by several order of magnitudes. In this way, the full human genome can be
analyzed, opening a unique possibility to understand the existence and
nature of the correlations between thermally activated bubbles and promoters.  

\section {Acknowledgments}
Work at Los Alamos National Laboratory is supported by the US Department of Energy under contract 
contract No. W-7405-ENG-36.


\begin{thebibliography}{10}
\bibitem{para}The parameters were chosen in Ref. \cite{campa} to fit thermodynamic properties of DNA: 
$k=0.025eV/A^2$, $\rho=2$, $\beta=0.35A^{-1}$ for the inter-site coupling;
for the Morse potential $D_{GC}=0.075eV$, $a_{GC}=6.9A^{-1}$ for a G-C
bp, $D_{AT}=0.05eV$, $a_{AT}=4.2A^{-1}$ for the A-T bp.
\end{thebibliography}
\end{document}